# Scalable quantum random number generator for cryptography based on the random flip-flop approach

Mario Stipčević[1]*, Ivan Michel Antolović[2,3], Claudio Bruschini[2], Edoardo Charbon[2]

[1]Ruđer Bošković Institute, Bijenička 54, 10000 Zagreb, Croatia

[2]Ecole polytechnique fédérale de Lausanne (EPFL), 2002 Neuchâtel, Switzerland

[3]Pi Imaging Technology SA, Rue de la Pierre-â-Mazel 39, 2000 Neuchâtel, Switzerland

*E-mail: stipcevi@irb.hr

**For globally connected devices like smart phones, personal computers and Internet-of-things devices, the ability to generate random numbers is essential for execution of cryptographic protocols responsible for information security. Generally, a random number generator should be small, robust, utilize as few hardware and energy resources as possible, yet provide excellent randomness at a high enough speed (bitrate) for a given purpose. In this work we present a quantum random number generator (QRNG) which makes use of a photoelectric effect in single-photon avalanche diodes (SPADs) as a source of randomness and is scalable to any desired bitrate. We use the random flip-flop method in which random bits are obtained by periodic sampling of a randomly toggling flip-flop. For the first time we investigate this method in detail and find that, out of two main imperfections, bias is due only to hardware imperfections while autocorrelation predominantly resides with the method itself. SPADs are integrated on a silicon chip together with passive quenching and digital pulse-shaping circuitry, using a standard 180 nm CMOS process. A separate FPGA chip derives random numbers from the detection signals. The basic QRNG cell, made of only two SPADs and a few logic circuits, can generate up to 20 Mbit/s that pass NIST statistical tests without any further postprocessing. This technology allows integration of a QRNG on a single silicon chip using readily available industrial processes.**

**Introduction**

In an age when information is the most valuable commodity, the science and art of information protection - cryptography - has become ubiquitous and used by virtually anyone on the planet, several times a day. On each log into a mobile phone or a computer, access to e-mail via web, payment via a smart card, e-banking or mobile banking, cryptography is at work to ensure authenticity, secrecy, privacy, nonrepudiation and integrity of digital communications and information. To that end, random numbers are needed. Namely, random numbers are essential for cryptographic protocols by being the only part of a protocol that is not known in advance. It turns out that pseudo-randomness, being deterministic, cannot substitute true randomness. A pseudo-random number generator merely delegates the task of generating randomness to its seed value, which again must be obtained by a physical random number generator (RNG), and yet because it is short, the seed presents a significant vulnerability. Therefore, a physical RNG is a crucial ingredient of secure cryptography.

Random numbers are aslo indispensable in modeling and Monte Carlo simulations [27], randomized algorithms [33] and intensely researched biomimetic computation paradigms relevant to artificial intelligence such as spiky neurons [7] and random pulse computing [8], [29]. Examples of usage of random numbers in cryptography include: private key in the RSA asymmetric key pair, Diffie-Hellman protocol, Solovay-Strassen primality testing, challenge-response data, *salt* for hashing and nonce [28].

Physical RNGs can be divided into two categories: *asynchronous* which produce random numbers at their own, random pace, and *clocked* which produce a random number synchronous to an internal or external periodic clock signal. In order to deliver random numbers to a clocked system, such as a computer, asynchronous RNGs require buffering or an accumulation basin for numbers, while still leaving a finite probability of not having a number available at the time of a request. It is also non-trivial to combine two or more such RNGs into one, faster RNG. Another challenge is to ensure that the basin of random numbers cannot be tampered with or accessed from an attacker. All this complicates and limits usability of asynchronous RNGs. Examples include RNGs which use: photon arrival time [1], von Neumann or Peres extractor [2], quantum entanglement [3], etc. In contrast, clocked RNGs guarantee: 1) availability of a fresh random number at the time of request, and 2) allow for easy parallelization of an arbitrary number of RNGs in order to achieve any desired speed. Both feats are achievable directly, without buffering. Examples of clocked RNG methods include: arrival/no-arrival of photon within a constant period of time [2], logic state set/reset action [4] or arbiter-based generation [5], etc.

In this work we present and study a simple RNG, shown in Fig. 1, based upon a periodic triggering of a random flip-flop (RFF), a theoretical concept of a novel logic circuits introduced in our previous work [6]. In a nutshell, a RFF of a certain type (eg. T-type RFF, shown in Fig. 1) behaves just as the usual flip-flop of the same type, with the only exception that its clock input (denoted CP) does not act with certainty but randomly with a probability of one-half. In this study we will only need the "toggle" or "T-type" RFF, denoted TRFF for short. Because the toggling probability is one-half by design, the initial state of the flip-flop is irrelevant: the next state will either be 0 or 1, with equal probability, and therefore a TRFF can serve as a random number generator. It is easy to generate more, say $N$, random bits at once: the same clock signal may be shared by $N$ TRFFs resulting in generation of an *N*-bit word per each clock. Bits are generated at a clock frequency $f_{BIT}$. If a faster-paced bit generation is needed, a higher clock frequency can be used or several clocks mutually shifted in phase can be distributed to several TRFFs or groups of TRFFs. By definition, every TRFF acts as an independent random number generator that can work synchronously or asynchronously to others - possibilities of combining them are virtually endless. Notably, applications that may profit from plurality of independent RNGs are the aforementioned spiky neurons and stochastic computing.

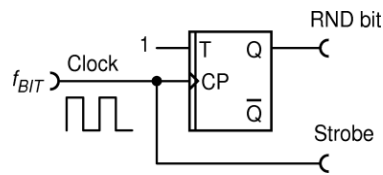

**Figure 1**. T-type random flip-flop (TRFF) connected as a random number generator: upon each clock pulse, a fresh random bit is generated synchronously.

However, the question is how to realize a TRFF in practice, while bearing in mind that scientific provability of its randomness is necessary for most applications, notably cryptographic ones, and that most real-life applications prefer a cheap, miniature, low-power device, while possibly requiring a large number of them. All these requirements drive our attention towards the use of intrinsic quantum randomness and technology of on-chip integration.

In recent years we witness a tremendous development of ultra-sensitive camera chips in which each pixel is a standalone single-photon detector comprising a single-photon avalanche photodiode (SPAD) operating in either free-running mode using passive [9], [24], or active [24] quenching, or in gated mode [25]. On top of that, each pixel may be equipped with a time-to-digital device (TDC) with time resolution on the order 10-100 ps [23], [10]. Chips carrying from tens up to a million single-photon pixels have been demonstrated using standard CMOS process. Furthermore, avalanche light sources [9-11], light emitting diodes (LED) [12-13] and light guides as well as light conduits [14] have been integrated too using industry standard CMOS process. These achievements open a possibility to realize multiple independent RFFs on a chip. In this work we investigate this path.

The paper is organized as follows. First, we first propose an experimental realization of a T-type RFF whose random clock action is derived from randomness of a quantum process of photoelectric effect and use it to build a quantum random number generator (QRNG). We then develop a theoretical model of main imperfections that cause this generator to generate non-perfect random numbers. With insight gained by studying that model, we build an improved RFF and use it for generation of long sequences of random numbers. Generated numbers are tested with NIST Statistical Test Suite (STS) version 2.1.2. [15].

**A practical realization of TRFF (methods)**

Building upon our earlier work on a general RFF, which consists of 3 flip-flops and a generator of random Poisson events of quantum origin [1], [6], here we propose a simplified TRFF shown in Fig. 2, which requires only 2 flip-flops and is well suited for the purpose of generating of random numbers. Random bits are obtained by a DFF which periodically samples a TFF that toggles at random times.

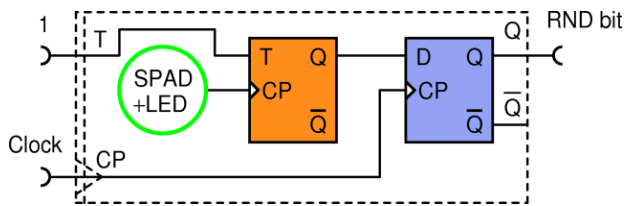

**Figure 2**. A practical realization of T-type random flip-flop whose random action relies on intrinsic quantum randomness of a photoelectric process.

In our experimental setup, we use a fully integrated SPAD detector array with passive quenching and recharge, described in our previous work [24], [26]. The chip, named SPAD23, has a physical size of 1.3 mm x 1.3 and contains 23 identical independent detectors (pixels) arranged in a honeycomb structure, shown in Fig. 3a, which fits within an area of about 0.115 mm x 0.115 mm in size. The area is homogeneously illuminated with an LED (Hamamatsu L7868, $\lambda$=670 nm, $\Delta\lambda$=30 nm FWHM) whose intensity may be set via an automated computer program so that pixels can deliver any desired count rate in range from 100 cps - 80 Mcps, where the low limit is set by dark counts. Since LED is a direct bandgap device, it emits photons at random times via quantum effect of spontaneous emission. On the other hand detection in a SPAD is a binomial process with a fixed probability, thus resulting detections are a random Poisson process. Because of this characteristic, LED illuminated photon detectors are frequently used in QRNGs [1],[4],[9],[30],[31]. We operate pixels at a low excess voltage (1V) which favors low afterpulsing, short dead time, high count rate and a low power consumption.

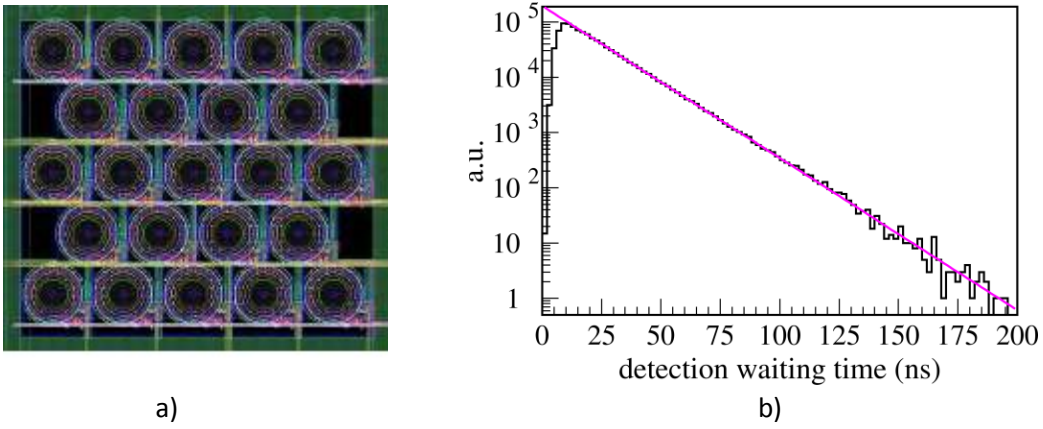

a)  b)

**Figure 3**. Micrograph of SPAD23 chip (a); distribution of waiting times of detection events from a single SPAD23 pixel illuminated with an LED at a count rate of 45 Mcps.

Figure 3b shows distribution of waiting times (times between subsequent detections) at a detection rate of 45 Mcps, as measured with a time-tagger (IdQuantique model ID900). The exponential fit (straight line) indicates the exponential probability density function (p.d.f.), characteristic of a Poisson process. Photon detectors exhibit dead time: the smallest time between two subsequent detections. The decline of the empirical distribution close to the zero waiting time is manifestation of the dead time of the pixel. Apparent non constancy of the dead time is a consequence of the passive quenching. The weighted average of the dead time is about $\tau_{DEAD} \approx 6$ ns. Estimation of aftterpulsing, using the method of Ref. [34], does not reveal any afterpulsing within error margin of 0.02%. Because of the very small afterpulsing probability, we omit afterpulsing from further analysis.

The TRFF is realized within a field-programmable gate array (FPGA) that is receiving detection impulses from the SPAD23 chip. For this study we use Intel-Altera Cyclone IV FPGA chip implemented in Terasic DE0-Nano board. In what follows, we model, analyze and quantify both technical and systemic imperfections of such a source of random numbers.

**Imperfection analysis: bias**

Assuming that photon detection is a stationary process, the QRNG shown in Fig. 2 will have a well defined probability $p_0$ of generating 0 and probability $p_1 = 1 - p_0$ of generating 1. Bias is defined as a deviation of $p_1$ from the ideal value of one-half:

$$b = p_1 - \frac{1}{2} = \left(\frac{1}{N}\sum_{i=1}^{N} x_i\right) - \frac{1}{2} \qquad (1)$$

where $x_i$ are individual bits in a bit string of $N$ bits. Having in mind that generation is a Binomial process with $N$ trials, the statistical variance of bias is: $\sigma_b = 1/(4N)$. Due to the toggling action of the T-type flip-flop (TFF), in theory the two probabilities are perfectly balanced and bias is zero. The idea of removing the bias by periodic sampling of the output state of a flip-flop toggling upon Poissonian random events was first described by C. H. Vincent in 1970 [16], generalized by Chevalier & Menard in 1974 [32] and independently re-discovered and used subsequently [17], [18], [19], [31]. However, in an experimental realization, unavoidable departures from this ideal picture lead to a non-zero bias - an effect that has not been studied before.

To model the bias we propose a simple model, illustrated in Fig. 4, which accounts for major imperfections that are most likely to appear in a practical realization of the RFF circuit shown in Fig. 2.

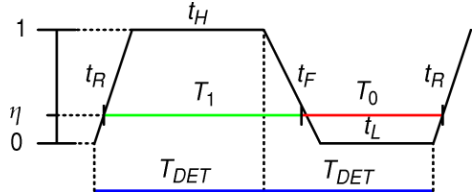

**Figure 4**. Waveform at the output Q of a toggling T-type flip-flop (black curve) and its logic value seen by a D-type flip-flop whose input D threshold level $\eta$ is somewhere between the LOW and HIGH levels (green and red lines).

In the model, we introduce several figures of merit. Regarding output Q of the TFF we define: rise time $t_R$, fall time $t_F$, duration of HIGH state $t_H$, and duration of LOW state $t_L$. Regarding DFF, we define threshold level $\eta \in [0,1]$ as a fraction of the voltage window between LOW and HIGH states at which D input triggers. Waiting time between two photon detections is an exponential random variable with mean $T_{DET}$. In order to model the average probability of ones, we consider two consecutive waiting times equal to $T_{DET}$. Duration of the average time $T_1$, during which DFF sees a HIGH state at its D input, is

$$T_1 = (1 - \eta)(t_R + t_F) + t_H \tag{2}$$

similarly, the average time $T_0$, during which DFF sees a LOW state at its D input, is

$$T_0 = \eta(t_R + t_F) + t_L. \tag{3}$$

From Fig. 4 we see that the following holds:

$$T_0 + T_1 = 2T_{DET} \tag{4}$$

$$t_H = T_{DET} - t_R \tag{5}$$

$$t_L = T_{DET} - t_F. \tag{6}$$

Probability of ones, generated by the DFF, is then equal to

$$p_1 = \frac{T_1}{T_0 + T_1} = \frac{1}{2} + \frac{t_F - \eta(t_R + t_F)}{2T_{DET}}. \tag{7}$$

Substituting this to Eq. (1) and defining the SPAD average detection rate $f_{DET} = 1/T_{DET}$, we obtain the following expression for the bias:

$$b = \frac{t_F - \eta(t_R + t_F)}{2} f_{DET} = \alpha f_{DET}. \tag{8}$$

For a given FPGA and operating conditions, the parameters $\eta$, $t_R$, and $t_F$ are constant and therefore bias should be strictly proportional to $f_{DET}$ and independent on bit sampling frequency. This is indeed confirmed by measurements of bias as a function of detection rate for a set of bit sampling frequencies, shown in Fig. 5. From this we estimate $\alpha \approx 6.8$ ps for the FPGA we used.

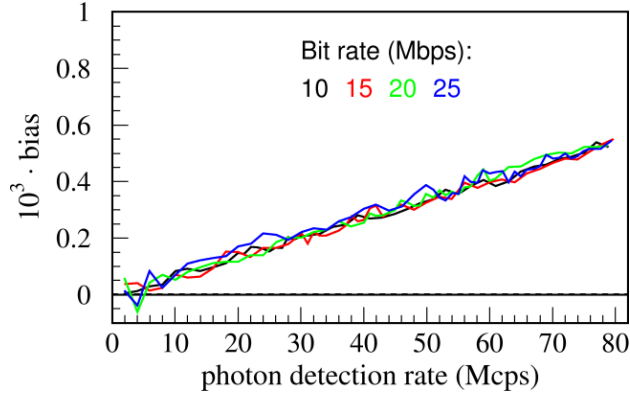

**Figure 5**. Bias of random bits generated by the RFF (shown in Fig. 2) as a function of the photon detection rate, for Altera-Intel Cyclone IV chip. Bits were generated at a rate $f_{BIT} = 10, 15, 20$ and 25 MHz. As expected from the model, bias does not depend on the bit rate.

By setting $b = 0$ in Eq. (8) we get

$$\eta = \frac{1}{1 + t_R/t_F}. \tag{9}$$

For any $t_R > 0$, $t_F > 0$ there is therefore an $\eta \, \epsilon (0,1)$ value for which bias vanishes, regardless of the detection rate or bit sampling rate. However, in our setup $\eta$, $t_R$, and $t_F$ are constant parameters of the chip and cannot be adjusted. Regarding manufacturing precision required to achieve the observed level of bias, by inserting $k$ into Eq. (8) one obtains $t_R - t_F \approx 28$ ps, while taking into account datasheet value of $t_R, t_F \sim 500$ ps one concludes that $0.486 \leq \eta \leq 0.514$. With tolerances so tight, no substantial improvement of bias can be expected by improving the FPGA manufacturing: in fact, the attained precision level is already quite amazing. Since several orders of magnitude lower bias is required for most applications, as well as to pass relevant statistical tests, we will show another way to improve it.

**Imperfection analysis: autocorrelation**

When a QRNG, based upon a TRFF circuit shown in Fig. 2, is optically and electrically well isolated from the environment, there will be no correlations between a bit it generates and any external information. However, empirically we find a substantial autocorrelation, defined through serial autocorrelation coefficients of lag $k$ [19]

$$a_k = \frac{\sum_{i=1}^{N-k}(x_i - \bar{x})(x_{i+k} - \bar{x})}{\sum_{i=1}^{N-k}(x_i - \bar{x})^2} \tag{10}$$

where $x_i$ is *i*-th bit in the generated bit stream. Coefficients $a_k$ are normalized to interval [-1, 1] with $-1$ being a maximum anti-correlation (toggling bit values) and $+1$ being a maximum correlation (identical bit values). Given the length $N$ of the bit stream, Eq. 10 gives an estimate of a coefficient with statistical variance of $1/(N-k-1) \approx 1/N$. The interpretation of $a_k$ is as follows. Given value of bit $x_i$, probability $s_k$ that $x_{i+k}$ has the same value is

$$s_k = 1/2 + a_k/2. \tag{11}$$

If the stream is truly random, then $a_k \equiv 0$.

To the best of our knowledge, autocorrelation of circuit in Fig. 2 was not studied so far either. To understand origins of autocorrelation we model the largest (leading) autocorrelation coefficient $a_1$ for the circuit in Fig. 2 in a simple case in which the SPAD dead time is negligible. Probability that exactly $k$ detections will appear between two samples is given by Poisson distribution:

$$P(k, \lambda) = e^{-\lambda} \frac{\lambda^k}{k!} \tag{12}$$

where $\lambda = f_{DET}/f_{BIT}$ is the normalized event rate. Effectively, $\lambda$ is a mean number of random events (photon detections) per generated bit. Let us imagine the TFF is sampled (by the DFF) and a certain bit value is generated. The next bit will have the same value if and only if an even number (0, 2, 4,...) of detections happens during the readout period. Probability of having the same bit value after one sample period is equal to the sum of probabilities of having even number of detections in between:

$$s_1 = \sum_{k=0}^{\infty} P(2k, \lambda) = e^{-\lambda} \sum_{k=0}^{\infty} \frac{\lambda^{2k}}{2k!} = e^{-\lambda}\cosh(\lambda) = \frac{1}{2} + \frac{1}{2}e^{-2\lambda}. \tag{13}$$

Inserting this into Eq. (11) we obtain the autocorrelation coefficient

$$a_1 = e^{-2f_{DET}/f_{BIT}}. \tag{14}$$

To verify this theoretical result, we generate bits at various $\lambda$ ratios and plot autocorrelation coefficient $a_1$ calculated via Eq. (10). Bits are obtained at a large sampling period of 4000 ns, so that the effect of the dead time (being about 6 ns) is negligible. Experimental results, shown in Fig. X, are in agreement with the model over several orders of magnitude of the autocorrelation coefficient.

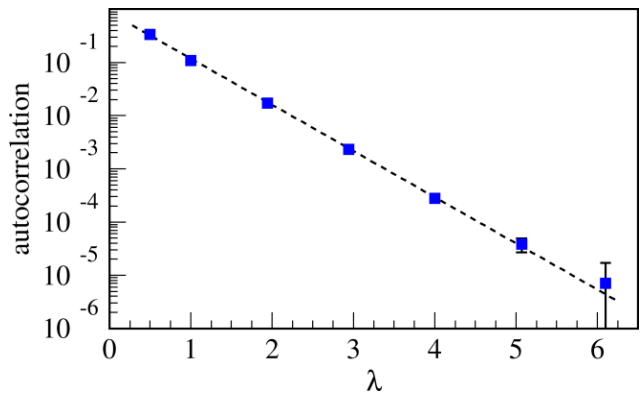

**Figure X**. Measured autocorrelation coefficients $a_1$ (blue squares with error bars) as a function of the ratio of photon detection and bit generation rates, $\lambda = f_{DET}/f_{BIT}$ (normalized event rate). Theoretical curve, according to Eq. (14), is represented by the dashed line.

The main departure from this simple model in a practical realization is that generally the SPAD dead time cannot be neglected. Measurements in Fig. 6 show the first four autocorrelation coefficients ($a_1$ to $a_4$) at significantly higher bit and photon generation rates. We note that the largest coefficient in amplitude, $a_1$, rises sharply as $\lambda \to 0$. This is because if the TFF is sampled too frequently with regard to its mean toggling rate, then there is an enhanced probability of obtaining the same bit twice, which is reflected as a positive autocorrelation. This effect accounts for autocorrelation behavior which tends to 1 as $f_{DET}/f_{BIT} \to 0$. However, situation is complicated by the dead time: if the sampling period $1/f_{BIT}$ is somewhat larger than the dead time, then there will be an enhanced probability that TFF will toggle its state exactly once between two samples, which implies anti-correlation. The effect is more pronounced as the sampling period approaches dead time or as detection rate rises. The interplay of the two effects, one causing correlation and the other causing anti-correlation, may create a sign-changing behavior of autocorrelation coefficients, such as seen for $f_{BIT} \geq 20$ Mcps. For lower $f_{BIT}$, dead time is too short to make a noticeable effect, thus autocorrelation dies off exponentially with $\lambda$, according to Eq. (14).

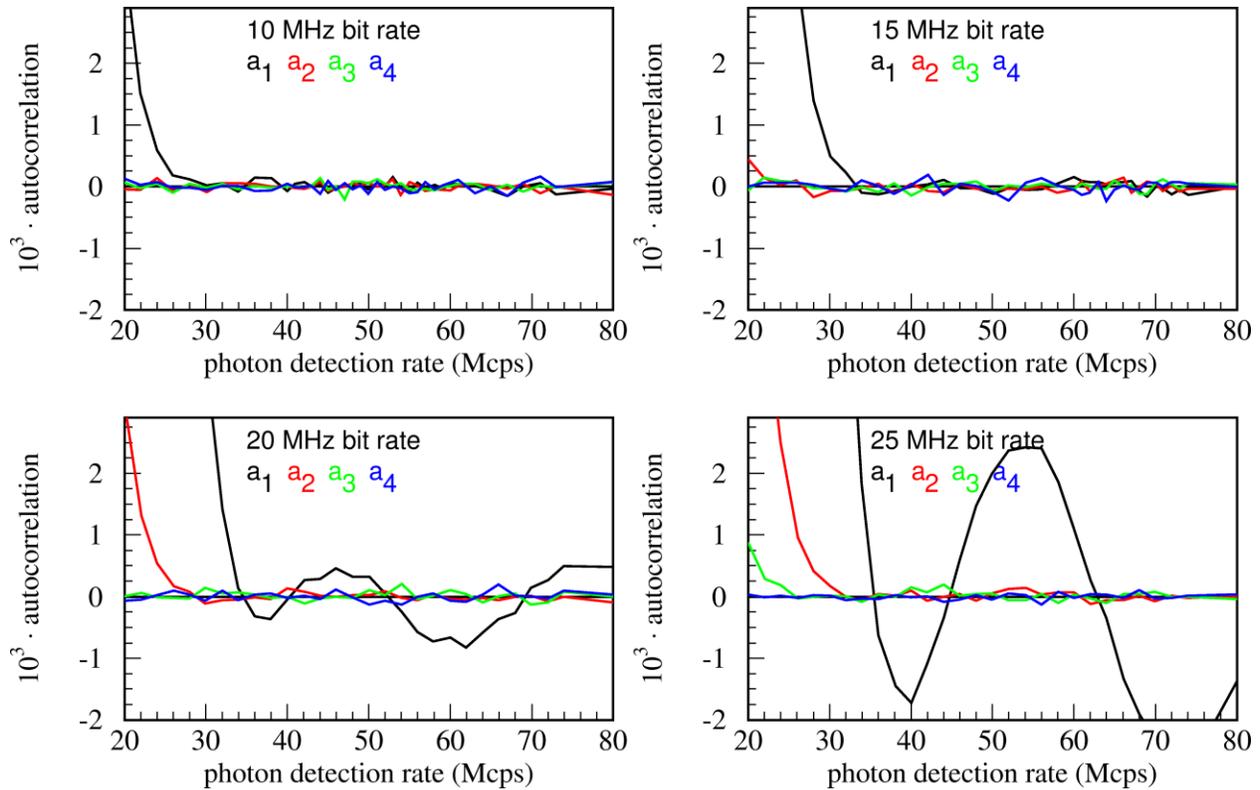

**Figure 6**. Autocorrelation coefficients $a_1$ through $a_4$ as a function of detector's detection rate for four bit sampling rates 10, 15, 20 and 25 MHz with $\tau_{DEAD} \approx 6$ ns. Colored curves show autocorrelation coefficients of different lags as a function of the photon detection rate.

The conclusion of this analysis is that even if an ideal circuit as in Fig. 2 could be fed by perfectly random events having zero dead time, it would still generate autocorrelated bits. The autocorrelation is intrinsic to this bit generation method and would be present even if hardware was perfect, in contrast to bias which is only present due to hardware imperfections. An empirical rule of thumb is that autocorrelation, for the circuit in Fig. 2, will be within $\pm 10^{-3}$ if both of the following conditions are met: $f_{DET} \geq 2.5 f_{BIT}$ and $\tau_{DEAD} \approx 1/(8 \cdot f_{BIT})$.

**Improved TRFF**

In order to improve on both the bias and the autocorrelation of generated bits, one could use extractors [20] but that would ruin the simplicity of RNG being a single logic circuit: the RFF, which returns a random bit upon a request, along with other aforementioned advantages of this approach. Therefore, we resort instead to constructing a better RFF. In doing so we set three requirements: 1) to improve precision without too high a hardware cost; 2) to preserve low clock-to-bit latency; and 3) to have ability to set an upper estimate of randomness imperfection of the improved RFF. To accomplish all requirements, the simplest solution is XOR-ing of two single-stage circuits shown in Fig. 2. This, in effect, results in an improved, double-stage TRFF, shown in Fig. 7. Having in mind the intended VLSI integration of RFF, this is an acceptable approach in terms of hardware cost. The latencies of single and double-stage TRFF, in an FPGA design, are the same.

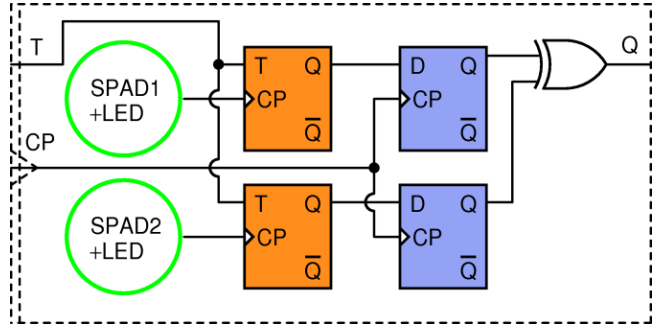

**Figure 7**. Realization of an improved, double-stage TRFF.

According to [21] XORing two statistically independent random strings, each with bias $b$ and autocorrelation $a_1$, results in a new string with an improved bias $b'$ and autocorrelation $a_1'$:

$$b' = -2b^2 \qquad (11)$$

$$a_1' = a_1^2 + 8a_1 b^2 \qquad (12)$$

If we chose as working condition $f_{BIT} \leq 20$ MHz and set $f_{DET}$ anywhere in the broad range $35\ldots55$ Mcps, from Figs. 5 and 6 we see that: $b \leq 5 \cdot 10^{-4}$; $|a_1| \leq 5 \cdot 10^{-4}$. By applying Eqs. (11) and (12), we estimate the upper bounds for bias and autocorrelation of the improved double-stage RFF: $|b'| \leq 5 \cdot 10^{-7}$ and $|a_1'| \leq 2.5 \cdot 10^{-7}$. Taking into account that statistical error of estimation of the bias and correlation of an $N$ bits long sequence is on the order of $1/\sqrt{N}$, we conclude that an XOR-ed sequence shorter than $N \sim 10^{13}$ bits will be de facto indistinguishable from a truly random sequence by *any* statistical test. One could further improve the RFF by XOR-ing more than two single-stage RFFs, while chaining Eqs. (11) and (12) would yield upper bound estimates for the multi-stage RFF.

Since the improved TRFF uses two (or potentially more) SPADs which are likely to be close to each other on the chip, there could be an optical or electrical crosstalk that would cause all involved TFFs to toggle (almost) simultaneously. However, unconditional toggling of all TFFs does not create a correlation and on average it does not change bias, thus randomness of the composite TRNG is the same as if there was no crosstalk. Therefore, crosstalk does not need to be included in our model.

**Testing of TRFF-based QRBG**

We build QRBGs based on single-stage and double-stage TRFF. Referring to Fig. 6, we chose a demanding operating point $f_{BIT} = 20$ Mbit/s, $f_{DET} = 45$ Mcps, at which bias and correlation of the single-stage RFF are not the best possible, in order to study effectiveness of the improvement achieved by adding the second stage to TRFF.

To enhance sensitivity of statistical testing, we use 25 strings of $10^9$ bits generated by each QRBG (single-stage and double-stage). We obtain the following median bias and autocorrelation of the 25 strings: $b = (250 \pm 16) \cdot 10^{-6}, a_1 = (299 \pm 32) \cdot 10^{-6}$ for the single-stage RFF and $b = (-5 \pm 6) \cdot 10^{-6}, a_1 = (= 3 \pm 12) \cdot 10^{-6}$ for the double-stage TRFF. As expected, bias and autocorrelation are indeed much lower for the double-stage RFF and are close to predictions of Eqs. (11) and (12) given measurements of the single-stage TRFF.

Next, we test both generators using the NIST STS, but with a twist of combining test results of 25 independent strings instead of reporting only one, as done by vast majority of previous art.

Empirically, we note that most of the tests in the STS battery are passed every time even by the single-stage TRFF. These tests largely belong to Marsaglia's diehard battery which is devoted to detecting complicated multi-dimensional correlations normally present in pseudo-random sequences, but highly improbable in our quantum-physical generator. We therefore turn our attention to the four tests that seem to have more relevance in our case. The frequency test counts zeros and ones and is clearly the most sensitive test to bias; Shannon entropy is affected both by bias and autocorrelation because they make a difference in probabilities of different n-grams (for a true random string those probabilities should all be the same); fast discrete Fourier transform test (FFT) would tell us if there are periodic frequencies hidden in our Poisson source; while Maurer's universal test checks for cryptographic weaknesses [22].

The way NIST STS works on a string of bits is that it runs each test on all available consecutive non-overlapping blocks of $10^6$ bits (the default block size) thus gathering statistics to form a p-value for the test. But a single p-value does not tell us much unless it is very close to zero (e.g. < 0.001), in which case it tells us that a particular test has failed with a high probability. However, if the test is run on multiple independent and truly random samples, the p-value should be uniformly distributed between 0 and 1. A practical way to visualize a distribution of p-values is to sort them in ascending order and plot as a function of the ordinal number. The resulting cumulative distribution function (c.d.f) is monotonic and for uniform distribution it should rise linearly rise from 0 to 1, up to the statistical variation. A test failure is indicated by an excess of small p-values.

Experimental results obtained for the single-stage TRFF are shown in Fig. 8 left. We see that, due to the clearly measurable bias, the frequency test is failed far too often. On the other hand, the entropy test seems to be passed quite well (blue dashed line), but this is only because the default n-gram size $L$ is set to 10 bits. There are $2^L$ different n-grams and within the default block size of $10^6$ bits each n-gram appears about 976 times - too few to statistically detect effects of the tiny bias and autocorrelation.

Besides that, we have seen that autocorrelation in our generator is of a very short range - only two neighboring bits are slightly correlated - thus the block length of 10 is far from optimal. To improve significantly on both aspects, we decrease the n-gram length to $L = 3$ bits.

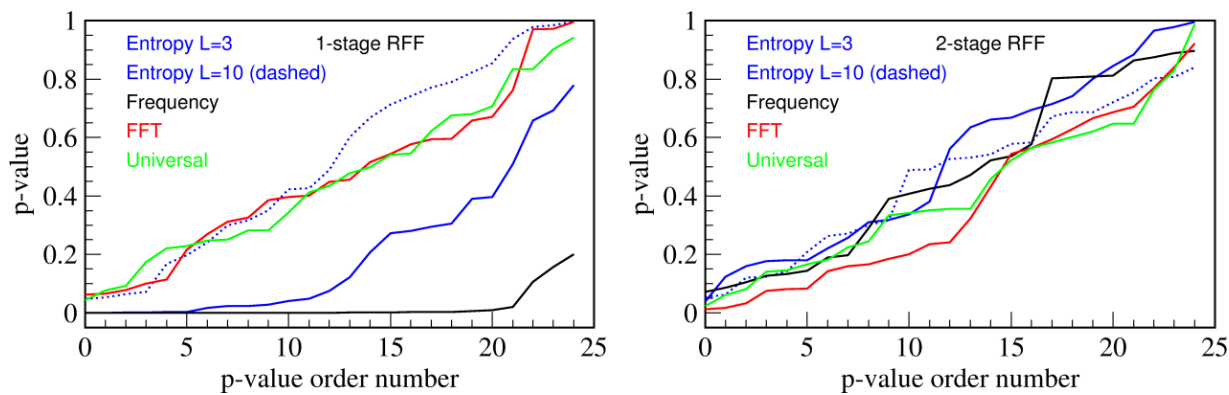

**Figure 8**. Entropy (block lengths 3 and 10 bits), frequency, fast Fourier, and Maurer's Universal test p-values, for 25 generated sequences, each $10^9$ bits long, sorted in ascending order. A failure is indicated by an excess of small p-values, as seen for Frequency and Entropy (L=3) tests for the 1-stage TRFF.

The Entropy test is failed too often (blue line). Maurer's Universal and FFT tests are passed as well as all other tests in the NIST STS suite indicating good quality of the generator in their respect. We repeat the same set of tests for the double-stage TRFF and find that it passes them all, as shown in Fig. 8 right.

| **Statistical test** | ***p*-value** | **Proportion/Threshold** | **Result** |
|---|---|---|---|
| Frequency | 0.442831 | 989/980 | PASS |
| BlockFrequency | 0.378705 | 990/980 | PASS |
| CumulativeSums | 0.560006 | 987/980 | PASS |
| Runs | 0.251837 | 989/980 | PASS |
| LongestRun | 0.637119 | 988/980 | PASS |
| Rank | 0.925287 | 989/980 | PASS |
| FFT | 0.397688 | 988/980 | PASS |
| NonOverlappingTemplate | 0.521158 | 990/980 | PASS |
| OverlappingTemplate | 0.233162 | 989/980 | PASS |
| Universal | 0.285427 | 985/980 | PASS |
| ApproximateEntropy | 0.388990 | 993/980 | PASS |
| RandomExcursions | 0.448798 | 611/604 | PASS |
| RandomExcursionsVariant | 0.587997 | 613/604 | PASS |
| Serial | 0.530182 | 990/980 | PASS |
| LinearComplexity | 0.099513 | 990/980 | PASS |

Table 1. Typical results of NIST statistical test suite STS-2.1.1 for 1000 samples of $10^6$ bits generated with the double-stage TRFF QRNG / XORed QRNG. For each statistical test an overall *p*-value as well as proportion of samples that passed the test versus theoretical threshold are given.

Finally, we apply the full NIST STS battery to test the 25 strings generated by the double-stage TRFF QRNG and find that all strings pass. We use plain data directly coming out of the device without any post-processing. A typical result of a test is shown in Table 1.

**Conclusion**

A conceptually simple quantum random number generator consists of a single electronic circuit - the random flip-flop - which returns one random bit upon a request. Operation of the T-type random flip-flop (TRFF) presented here is based on a known principle of sampling a logic state of a flip-flop toggling upon a discrete random process, in this case single-photon detection through the photoelectric effect in silicon avalanche photodiodes. However, for the first time a theory of operation of such a circuit is developed here, which allowed us to define and analyze main sources of its non-randomness, namely bias and serial autocorrelation. We have shown that the bias is caused exclusively by technological imperfections or limitations such as finite state-transition times and an unbalanced threshold level of flip-flops that constitute a TRFF, while autocorrelation is intrinsic to the method of generating random numbers and would persist even if flip-flops would be free of any imperfections. Furthermore, we find that in case of non-vanishing dead time, autocorrelation is modified in a complex interplay of bit generation rate, detection rate and dead time, in which the dead time can either improve or worsen the autocorrelation.

This analysis helped us to build an improved TRFF capable of generating random numbers that pass NIST STS statistical test suite at a generation rate of 20 Mbit/s, while still being reasonably simple and "cheap" in terms of hardware requirements.

Because a TRFF generates random bits in synchronization to a clock, it is easy to parallelize or otherwise combine operation a number of TRFFs in order to generate random bits at any desired rate.

Finally, we conjecture that a TRFF based QRNG might be offering better security than extractor-based ones which must store a large block of bits internally, because no bit is stored in the generator: there is a direct, memoryless path between the quantum-random process and an output bit. We believe that TRFF, as a uniquely non-deterministic logic element, would be an invaluable addition to future FPGAs and cryptographic chips, as a natural and simple-to-use source of randomness.

**References**

1. Stipčević M., Medved Rogina B., Quantum random number generator based on photonic emission in semiconductors, *Rev. Sci. Instrum.* **78**, 045104:1-7 (2007). DOI: [10.1063/1.2720728](10.1063/1.2720728)
2. A. Stanco, D. G. Marangon, G. Vallone, S. Burri, E. Charbon, P. Villoresi, "Efficient random number generation techniques for CMOS SPAD array based devices", Phys. Rev. Research **2**, 023287 (2020) DOI: 10.1103/PhysRevResearch.2.023287.
3. G. Vallone, D. G. Marangon, M. Tomasin, and P. Villoresi, "Quantum randomness certified by the uncertainty principle", Phys. Rev. A 90,052327(2014). DOI: 10.1103/PhysRevA.90.052327.
4. T. Jennewein, U. Achleitner, G. Weihs, H. Weinfurter, A. Zeilinger, "A Fast and Compact Quantum Random Number Generator", Rev. Sci. Instrum. **71**, 1675-1680 (2000).
5. N. Massari et al., "16.3 A 16×16 pixels SPAD-based 128-Mb/s quantum random number generator with −74dB light rejection ratio and −6.7ppm/°C bias sensitivity on temperature," 2016 IEEE ISSCC conference, San Francisco, CA, 2016, pp. 292-293. DOI: 10.1109/ISSCC.2016.7418022.

6. M. Stipčević, "Quantum random flip-flop and its applications in random frequency synthesis and true random number generation", *Rev. Sci. Instrum*. **87**, 035113 (2016). DOI: 10.1063/1.4943668.
7. W. Maass, "Networks of spiking neurons: The third generation of neural network models". *Neural Networks*. **10**, 1659–1671 (1997). DOI:10.1016/S0893-6080(97)00011-7.
8. A. Alaghi, W. Qian and J. P. Hayes, "The Promise and Challenge of Stochastic Computing," in *IEEE Transactions on Computer-Aided Design of Integrated Circuits and Systems*, vol. 37, no. 8, pp. 1515-1531, Aug. 2018, DOI: 10.1109/TCAD.2017.2778107.
9. F. Acerbi et al., "Structures and Methods for Fully-Integrated Quantum Random Number Generators", IEEE J. Sel. Top. Quantum Electron **26**, 1-8 (2020). DOI: 10.1109/JSTQE.2020.2990216.
10. N. Massari et al., "A Compact TDC-based Quantum Random Number Generator", 2019 26th IEEE International Conference on Electronics, Circuits and Systems (ICECS), Genoa, Italy, 2019, pp. 815-818, DOI: 10.1109/ICECS46596.2019.8964941.
11. A. L. Lacaita, F. Zappa, S. Bigliasrdi, and M. Manfredi, "On the Bremsstrahlung origin of hot-carrier-induced photons in silicon devices," *IEEE Trans. Electron Devices* **40**, 577–582 (1993).
12. J. F. C. Carreira, A. D. Griffiths, E. Xie, B. J. E. Guilhabert, J. Herrnsdorf, R. K. Henderson, E. Gu, M. J. Strain, and M. D. Dawson, "Direct integration of micro-LEDs and a SPAD detector on a silicon CMOS chip for data communications and time-of-flight ranging," *Opt. Express* 28, 6909-6917 (2020). DOI 10.1364/OE.384746.
13. K. Xu, B. Huang, K. A. Ogudo, L. W. Snyman, H. Chen, and G. P. Li, "Silicon Light-emitting Device in Standard CMOS technology," in International Photonics and OptoElectronics, OSA Technical Digest (online) (Optical Society of America, 2015), paper OT1C.3. DOI: 10.1364/OEDI.2015.OT1C.3.
14. Bisadi, Z., Meneghetti, A., Tomasi, A., Tengattini, A., Fontana, G., Pucker, G., Bettotti, P., Sala, M., & Pavesi, L. (2016). Generation of high quality random numbers via an all-silicon-based approach. Physica Status Solidi (a), 213, 3186-3193. DOI:10.1002/PSSA.201600298.
15. Rukhin A. *et al.*, NIST Special Publication 800-22rev1a (April 2010), URL: http://csrc.nist.gov/rng , Date of access: 01/02/2012.
16. C. H. Vincent, The generation of truly random binary numbers. J. Phys. E: Sci. Instrum. **3**, 594–598 (1970).
17. V. Bagini, M. Bucci, "A design of reliable true random number generator for cryptographic applications", in *Cryptographic Hardware and Embedded Systems (CHES)*, ed. by Ç.K. Koç, 1206 C. Paar (Springer, Berlin, 2002), pp. 204–218.
18. Stipčević M., Fast nondeterministic random bit generator based on weakly correlated physical events, *Rev. Sci. Instrum.* **75**, 4442-4449 (2004).
19. M. Stipčević, J. Bowers, "Spatio-temporal optical random number generator", Opt. Express **23**, 11619-11631 (2015). DOI: 10.1364/OE.23.011619.
20. Y. Dodis, "On extractors, error-correction and hiding all partial information," *IEEE Information Theory Workshop on Theory and Practice in Information-Theoretic Security, 2005.*, Awaji Island, 2005, pp. 74-79, DOI: 10.1109/ITWTPI.2005.1543961.
21. Davies R., *Exclusive OR (XOR) and hardware random number generators*, February 28, 2002, URL: http://www.robertnz.net/pdf/xor2.pdf, Date of access: 05/02/2014.
22. U. M. Maurer, "A universal statistical test for random bit generators.", J. Cryptology **5,** 89–105 (1992). DOI: 10.1007/BF00193563.
23. S. Burri, C. Bruschini, E. Charbon, "LinoSPAD: A Compact Linear SPAD Camera System with 64 FPGA-Based TDC Modules for Versatile 50 ps Resolution Time-Resolved Imaging", MDPI Instruments, **1**(1), 6 (2017). DOI: 10.3390/instruments1010006.
24. I. M. Antolovic, C. Bruschini, and E. Charbon, "Dynamic range extension for photon counting arrays," Opt. Express **26**, 22234-22248 (2018).
25. Ulku A, Ardelean A, Antolovic M, Weiss S, Charbon E, Bruschini C *et al.* Wide-field time-gated SPAD imager for phasor-based FLIM applications. *Methods Appl Fluoresc* 2020; **8**: 024002. DOI:10.1088/2050-6120/ab6ed7.
26. SPAD23, Pi Imaging Technology Sa., https://piimaging.com/datasheet/SPAD23system.pdf, last visited 17.02.2021.

27. B. Dang *et al*., "Physically Transient True Random Number Generators Based on Paired Threshold Switches Enabling Monte Carlo Method Applications," in *IEEE Electron Device Letters*, vol. 40, no. 7, pp. 1096-1099, July 2019, DOI: 10.1109/LED.2019.2919914.
28. R. Gennaro, "Randomness in Cryptography" in *IEEE Security & Privacy*, vol. 4, no. 02, pp. 64-67, 2006. DOI: 10.1109/MSP.2006.49
29. M. Batelić and M. Stipčević, "Improved circuits for a random pulse computer," *2020 43rd International Convention on Information, Communication and Electronic Technology (MIPRO)*, Opatija, Croatia, 2020, pp. 123-127, DOI: 10.23919/MIPRO48935.2020.9245116.
30. Xiaowen Li, Adam B. Cohen, Thomas E. Murphy, and Rajarshi Roy, "Scalable parallel physical random number generator based on a superluminescent LED," Opt. Lett. 36, 1020-1022 (2011). DOI: 10.1364/OL.36.001020.
31. H. Fürst, H. Weier, S. Nauerth, D. G. Marangon, C. Kurtsiefer, and H. Weinfurter, "", Opt. Express 18, 13029 (2010).
32. P. Chevalier, C. Menard, "Random number generator", U.S. Patent Number US3790768A, February 5, 1974.
33. R. Motwani, and P. Raghavan, ACM Computing Surveys 28, No. 1 (1996) DOI: 10.1145/234313.234327.
34. G. Humer, M. Peev, C. Schaeff, S., M. Stipčević, R. Ursin, "A simple and robust method for estimating afterpulsing in single photon detectors", J. Lightwave Technol. **33**, 3098-3107 (2015). DOI: 10.1109/JLT.2015.2428053.